\documentclass[aip, cha, reprint]{revtex4-2}

\usepackage[utf8]{inputenc}
\usepackage[T1]{fontenc}
\usepackage{mathtools}
\usepackage{graphicx}
\usepackage{hyperref}
\usepackage{etoolbox}  

\makeatletter
\def\@email#1#2{%
 \endgroup
 \patchcmd{\titleblock@produce}
  {\frontmatter@RRAPformat}
  {\frontmatter@RRAPformat{\produce@RRAP{*#1\href{mailto:#2}{#2}}}\frontmatter@RRAPformat}
  {}{}
}%
\makeatother

\begin{document}
\title[Link Cascades in Complex Networks]{Link Cascades in Complex Networks: A Mean-field Approach}
\date{\today}

\author{King Chun \surname{Wong}}
\affiliation{Department of Physics, Hong Kong University of Science and Technology, Clear Water Bay, Kowloon, Hong Kong, China}

\author{Sai-Ping \surname{Li}}
\affiliation{Department of Physics, Hong Kong University of Science and Technology, Clear Water Bay, Kowloon, Hong Kong, China}
\affiliation{Institute of Physics, Academia Sinica, Nankang, Taipei 115, Taiwan}

\email{spli@phys.sinica.edu.tw}

\begin{abstract}
	Cascade models on networks have been used extensively to study cascade failure in complex systems. However, most current models consider failure caused by node damage and neglect the possibility of link damage, which is relevant to transportation, social dynamics, biology, and medicine. In an attempt to generalize conventional cascade models to link damage, we propose a link cascade model based on the standard independent cascade model, which is then solved via both numerical simulation and analytic approximation. We find that the probability that a node loses all its links due to link damage exhibits a minimum as a function of node degree, indicating that there exists an optimal degree for a node to be most resistant to link damage. We apply our model to investigate the sign distribution in a real-world signed social network and find that such optimal degree does exist in real-world dataset.
\end{abstract}

\maketitle

\begin{quotation}
	Failure of complex networks often occur in a cascading manner, meaning that failure of one component of the system can trigger the failure of other nearby components, thereby causing a large-scale collapse of the entire system. While conventional cascade failure is studied by considering a damage that propagates from one node to another, the cause of failure can also be a damage in the links of the network. In this study, we propose a simple model to investigate the possibility and consequences of cascading link damage. Our results show that the vulnerability of a node under link damage is closely related to the number of links it is connected to. In particular, there is an optimal number of connections for a node to be most resistant to link damage. Our results provide insight to the design of complex networks and have potential applications in areas such as transportation, biology, and social science, where link damage is of concern.
\end{quotation}

\section{\label{sec:intro}Introduction}

	Complex networks have been used to model a wide range of dynamical processes in real-world complex systems, for example in biology \cite{avena-koenigsberger_brain, tang_brain}, finance \cite{elliott_financial}, social science \cite{castellano_social_dynamics}, and control theory \cite{liu_control}. Among these processes, cascade or diffusion processes describe how information, influence, or infection spread in a network and have many successful applications in epidemiology \cite{pastor-satorras_epidemic}, power engineering \cite{motter_power}, and the Internet \cite{pastor-satorras_internet}.
	
	The basic question in studying cascade processes is how a damage in one component of a system can propagate across the whole system. This is typically modeled by prescribing nodes in the network (representing components of the system) a state variable indicating the damage together with a state update rule that propagates the damage. For instance, in the susceptible–infectious–recovered (SIR) model \cite{newman_SIR} and the independent cascade (IC) model \cite{kempe_IC}, this is done by assigning nodes to be in either a susceptible, infectious, or removed state (also called silent, active, and inactive in the IC model). In Watt's threshold model \cite{watts_model}, nodes are assigned states of $0$ (off) or $1$ (on). Similarly, when interpreting site and bond percolation as cascade processes \cite{gleeson_cascades}, nodes can be in a state either inside or outside of the giant component. In all these common models as well as many other examples, the dynamics occurs between the nodes, and the links only serve as a bookkeeping device that keeps track of all possible interactions between nodes.

	\begin{figure*}[t]
		\includegraphics[width=\textwidth]{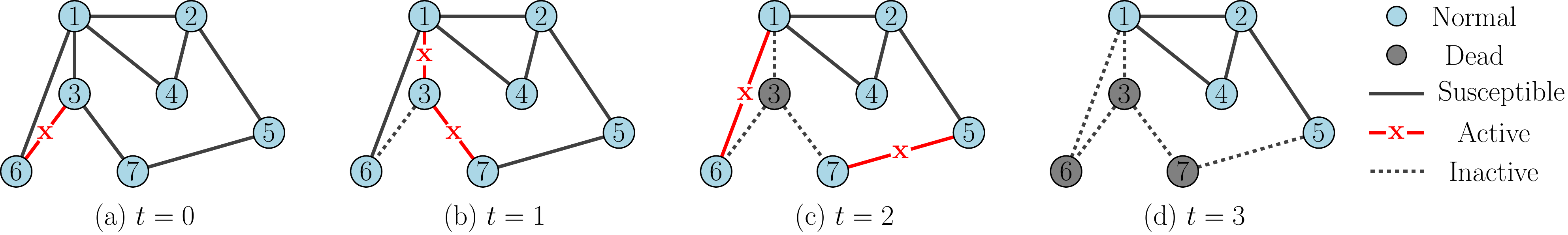}					
		\centering	
		\caption{\label{fig:lic-example} An illustrative example of our proposed model. (a) At $t=0$, a single seed link is chosen to start the cascade. (b) At $t=1$, the seed link successfully infects two more links. (c) At $t=2$, more infections occur. Node $3$ loses all its links and becomes dead. (d) At $t=3$, nodes $6$ and $7$ become dead also. The active links at $t=2$ fail to infect more links; there are no more active links to continue the spreading, and the cascade process terminates.}
	\end{figure*}
	
	In reality, failures can also be caused by damages in links without the explicit participation of nodes. As an example, in a transportation network, if we model geographical locations as nodes and their interconnecting roads as links, traffic congestion is then most suitably modeled as link damage. Congestion can be developed as results of accidents occurring on the road, and as congestion blocks the inflow traffic, more roads will become congested as well. Consequently, an initially broken link can trigger a cascade of link damages, while during the process, the nodes only passively witness the incident. Under this link cascade, a node may lose all its links and becomes effectively inaccessible and thus a failure point in the transportation network. This is known as the ``gridlock'' \cite{daganzo_gridlock} scenario in urban transportation. Another example of link cascade process can be realized in social networks. Let us suppose a conflict takes place between two friends and causes them to become enemies. Realizing their conflict, their acquaintances may take side on the issue and either remain as friends with or become enemies to them. This process may continue in a cascading manner, similar to opinion formation processes \cite{castellano_social_dynamics}. Since friendship and hostility are pairwise relations, they could naturally be modeled as a link property. Here again, the cascade process is solely driven by the dynamics of links instead of the nodes; conversely, it would be unrealistic to model social relation dynamics by having nodes to spontaneously turn into a villain state and become everyone’s enemy. Thus, the cascade of conflict described above would better be captured by a link-based process instead of a node-based process. Link cascade processes may also be observed in biological systems. One may consider the human circulation system as a network with blood vessels being the connecting links between tissues. Link damages occur when a disease or disorder, such as bacterial infection, cancer, and blood clots, causes a malfunction or blockage of blood vessels. As the disease or disorder spreads through the circulation system, the resulting cardiovascular obstruction similarly spreads, leading to a cascade of link damage. Here, the nodes again only passively perceive the link damage as the condition of blood flow changes and does not actively participate in the cascade. When such a link cascade damage occurs in the human brain, a stroke will result. In more extreme cases, when a node (tissue) loses all its links (blood vessels), the damage can result in a gangrene, leading to tissue death due to a lack of blood supply \cite{reid_pathology}.

	Despite the relevance of link cascades, to the best of our knowledge, Ref.~\onlinecite{feng_link_cascade} is the only recent work that explicitly investigated the notion of link cascades in a network. Ref.~\onlinecite{feng_link_cascade} proposed a link cascade model based on ``flow balance'' and studied the network's global response to link cascade by analyzing the giant connected component that remained after the cascade. Their model is however designed to mimic monetary flows in financial systems and therefore does not generalize well to other applications, such as the examples we have outlined above.
		
	In contrast to Ref.~\onlinecite{feng_link_cascade}, we consider a simple model that aims to represent more generic link cascade processes. Since the independent cascade model is one of the simplest possible representative node cascade models that had been studied with many applications \cite{jalili_IC, saito_IC, wang_IC, wong_IC}, we by analogy construct a ``link independent cascade'' model as one simplest representative link cascade model. We use this link independent cascade model as a prototype to explore features of link cascades in relation to network topology. Our key finding is that, under the link independent cascade model, the probability of node failure (defined as the probability a node lost all its links) is not a monotonic function of node degree $k$. In particular, there exists an optimal degree $k^*$ at which a node is least likely to fail. We demonstrate these findings using a combination of simulation evidence and analytic arguments. Our results have implications for the design and control of complex networks, especially in view of the many existing strategies for improving network robustness by link addition or removal (e.g. Ref.~\onlinecite{huang_robustness, zeng_robustness, wang_robustness, ji_robustness}) -- our results indicate that there is an optimal number of links that each node can have when link damage is of concern.	
	
	In Sec.~\ref{sec:model} we define our proposed model. In Sec.~\ref{sec:star}, we illustrate the features of our model by considering a star graph, in which our model is analytically tractable. In Sec.~\ref{sec:nwk}, we analyze our model in more generic network topology. In Sec.~\ref{sec:app}, we apply our model to the problem of friend-enemy network formation. We conclude our study in Sec.~\ref{sec:con}.

\section{\label{sec:model}Model Description}

	Let $\mathcal{G}$ be an undirected network of $N$ integer-labeled nodes $\{1, \dots, N \}$ interconnected by $E$ links. Two links are said to be ``neighboring'' if they share one same end node. Following the construction of the standard independent cascade model \cite{kempe_IC}, we assign to each link one of three states: susceptible, active, and inactive. The link states then evolve by the following dynamical rules together with a parameter (the ``diffusion probability'') $Q \leq 1$:

	\begin{enumerate}
		\item At time $t=0$, a fraction $\rho$ of links are chosen randomly to become active.
		\item At time $t>0$, each active link has a probability $Q$ to infect each of its susceptible neighboring links to become active at time $t+1$.
		\item After all possible infections have been attempted, active links become inactive. Inactive links cannot be infected again nor to infect others.
		\item If there are no more active links at $t+1$, terminate the process, otherwise forward to time $t+1$ and repeat the process from step 2.
	\end{enumerate}
	
	Therefore, a susceptible link is one that has not been infected, an active link is one that is infected and will actively spread the infection, and an inactive link is one that has been infected and removed from the network. Fig.~\ref{fig:lic-example} presents an illustrative example of the model. Simulation of our model can be straightforwardly performed using these defining rules, similar to the simulations in Ref.~\onlinecite{albert_error}.
	
	The assumption of ``independent cascade'' is accommodated in the infection step (Step 2); each attempt of infection is independent of all others with independent success probability $Q$. Note that under our dynamical rules, active links stay active for exactly one time-step, and therefore they can only make exactly one infection attempt per neighboring link. For simplicity, this paper will focus mainly on the case of a single seed link ($\rho = 1/E$), although our results can easily be generalized to arbitrary $\rho$ (Sec.~\ref{sec:nwk_4}).
	
	We are interested in studying how damage in links (representing inter-component connections) can lead to failure in nodes (representing the components themselves). To this end, we define a node to be \textit{dead} when all its links are inactive. The reason for this definition is that if the cascade is from link damages, a dead node would be one that has lost all its connections to the rest of the network. Hence a dead node is effectively removed and becomes a failure point in the network. In this paper, our goal is to study the survival statistics of nodes under link cascade in relation to the local network topology around the node.
	
	Before analyzing our model for general networks, we first study the simplest case of a star graph, which will serve as a basic building block in our analysis.

\section{\label{sec:star}Dynamics on Star Graph}
	\subsection{\label{sec:star_1}Markov Chain Formulation}
	Consider a star graph $\mathcal{S}_{k}$ in which the central node has $k$ neighbors. Despite its simplicity, the star graph is useful for studying dynamics at the single-node level: If we focus on a single node and cut off the network structure beyond its nearest neighbors, we will simply be left with $\mathcal{S}_{k}$, where $k$ is the degree of the focused node. The star graph results will provide us the basis to analyze the more general cases. 
	
	For the star graph, our model can be formulated exactly as a Markov chain. Let $n$ and $s$ respectively be the number of active and susceptible links that the central node has. The number of inactive links is therefore $r = k-n-s$ and is not an independent variable. Since there are no structures beyond the nearest neighbors, it makes no difference to distinguish which link is susceptible, active, or inactive. The state of the whole system can therefore be completely specified by the doublet $(n, s)$. With this parametrization, the dynamic rules defined in Sec.~\ref{sec:model} can be expressed mathematically by the following transition probabilities among the states:
	
	\begin{align}
		\label{eqn:star_trans}
		T_{star}\left(n, s \mid n', s' \right) = \delta_{s, s'-n} \binom{s'}{n} &\left(1 - (1-Q)^{n'} \right)^{n} \nonumber \\*
		&\times \left(1-Q \right)^{n's}.
	\end{align}

	\noindent The binomial form of $T_{star}$ follows directly from the independent cascade assumption: If the central node has $n'$ active link at time $t$, then each of its $s'$ susceptible links will experience $n'$ independent infection attempt, hence each susceptible link has probability $(1-Q)^{n'}$ to remain susceptible at $t+1$. The net transition would then be a binomial distribution as in Eq.~(\ref{eqn:star_trans}). The Kronecker delta is to ensure $s+n = s'$, so that all new susceptible and active links come from the old susceptible links. With this transition probability, the state distribution $P(n,s,t)$ evolves as
	
	\begin{equation}
		\label{eqn:star_recur}
		P\big(n, s, t+1) = \sum_{n', s'} T_{star}\big(n, s \mid n', s' \big) P\big(n', s', t),
	\end{equation} 
	
	\noindent subject to the initial condition $P(n, s, t=0) = \delta_{s, k-1} \delta_{n, 1}$, corresponding to a single seed link. Within this framework, a dead node is one with state $(n \geq 0, s=0)$. Since an active node becomes inactive in exactly one time step, all $n>0$ states are transient, and the dead probability can be defined as

	\begin{equation}
		\label{eqn:star_d}
		D(k) = \lim_{t\to \infty} P(n=0, s=0, t).
	\end{equation}
	
	The Markov chain defined by Eq.~(\ref{eqn:star_trans}) is known as the Reed-Frost binomial chain in mathematical epidemiology \cite{gani_jerwood_reed_frost}. The Reed-Frost chain admits no simple analytic solution, although some exact results can be obtained through the use of a special class of polynomials due to Gontcharoff \cite{lefevre_picard_reed_frost}. Here we will not follow this route of analysis, but instead we will directly perform the recursive sum Eq.~(\ref{eqn:star_recur}) numerically to obtain the dead probability $D(k)$ as a function of degree $k$.
	
	\begin{figure}[t]
		\includegraphics[width=\columnwidth]{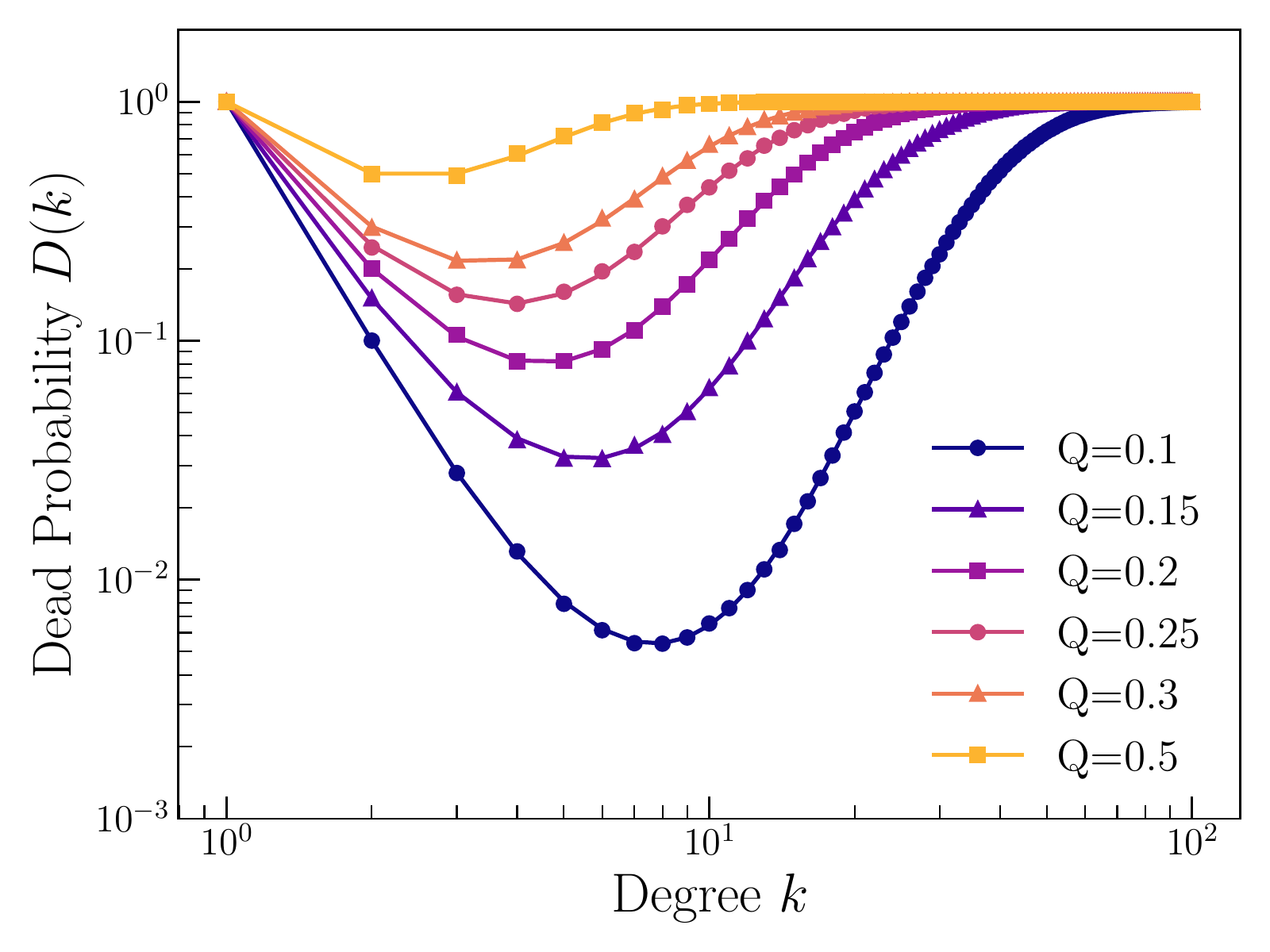}
		\centering	
		\caption{\label{fig:star} Dead probability $D(k)$ of a star graph $\mathcal{S}_k$, obtained from Eqs.~(\ref{eqn:star_trans})-(\ref{eqn:star_d}) (solid curves) and direct simulation of the dynamic rules in Sec.~\ref{sec:model} (markers). The results show exact agreements. The curves further indicate the existence of a minimum point $k_{star}^*$. In the case $Q=0.1$, while a node with $k\approx k_{star}^* = 8$ has $D \sim \mathcal{O}(10^{-3})$, a node with $k > 30$ has $D \sim \mathcal{O}(10^{-1})$. This is a $100$-fold difference in survivability.}
	\end{figure}

	Fig.~\ref{fig:star} shows the function $D(k)$ computed from Eqs.~(\ref{eqn:star_trans})-(\ref{eqn:star_d}) for $Q = 0.1, 0.15, 0.2, 0.25, 0.3, 0.5$ (solid curves). The same function is also measured from direct simulation of the dynamic rules defined in Sec.~\ref{sec:model} (markers). This is performed by counting the number of times the central node dies in an ensemble of simulation runs. The exact agreement between the two sets of values validates the Markov chain formulation. Moreover, all six curves show that $D(k)$ decreases for small $k$ and increases for large $k$. For all $Q$ values, $D(k)\to 1$ as $k \to \infty$. We see that for the star graph, the dead probability $D(k)$ always has a unique minimum at some optimal degree $k_{star}^*$. For example, for the case when $Q=0.1$, $D(k) \sim \mathcal{O}(10^{-3})$ around $k_{star}^*$, while $D(k) \sim \mathcal{O}(10^{-1})$ for $k>30$, giving a $100$-fold difference in survivability. Intuitively, one may expect that to protect a node from link failures, one should add more links to the node so that the unfortunate situation where all links fail at the same time becomes less likely. Our result however suggests the opposite -- an over-addition of links actually makes a node more vulnerable to link damage, and there exists an optimal number of links which keeps the node at its safest.

	\subsection{\label{sec:star_2}Optimal Degree}
	
	We can understand the origin of the optimal degree through a heuristic argument based on combinatorics. Let the observed states at time $t$ be $(n_t, s_t)$. $D(k)$ is then given by summing the realization probability of all possible sample paths that go from $(n_0=1, s_0 = k-1)$ to $(n_\infty = 0, s_\infty = 0)$, i.e. all possible ways to kill the node at the end of the cascade process. More precisely, it is the sum of realization probability over all possible infection sequences $(n_1, \dots, n_m)$ such that all $n_t > 0$ and $n_1 + \dots + n_m = (k-1)$, with $m \leq (k-1)$ being the length of the sequence. As an example, if $(k-1)=3$, the possible infection sequences would be $\{ 3 \}$, $\{1, 2\}$, $\{2, 1\}$, $\{1, 1, 1\}$, so that all $3$ initially susceptible links got infected in the corresponding $m=1, 2, 3$ steps. We may explicitly construct these probabilities in the $Q \ll 1$ limit. Expanding the transition probability (Eq.~(\ref{eqn:star_trans})) to lowest order in $Q$, we obtain
	
	\begin{equation}
		\label{eqn:star_trans_expanded}
		T_{star} = \delta_{s, s'-n} \binom{s'}{n} {n'}^{n}Q^n + \mathcal{O}(Q^{n+1}) \sim Q^n.
	\end{equation}
	
	\noindent This implies that we acquire a factor of $Q^{n_t}$ for each step of an infection sequence $(n_1, \dots, n_m)$. Since for all infection sequences, $\sum_t n_t = (k-1)$, the total realization probability for any of them is therefore $Q^{k-1}$. Next, we need to determine the total number of these infection sequences. Counting the number of ways to infect $(k-1)$ links in $m$ steps is combinatorially equivalent to distributing $(k-1)$ distinguishable balls into $m$ distinguishable bins such that each bin has at least one ball. From standard combinatorics, this is given by the Stirling number of the second kind multiplied by $m!$ \cite{rosen_combinatorics},

	\begin{equation}
		\label{eqn:stirling}
		m! \ S(k-1, m) = \sum_{i=0}^{m} (-1)^i \binom{m}{i}(m-i)^{k-1}.
	\end{equation}
	
	\noindent Summing over all possible step number $m$ produces the ordered Bell number or the Fubini number, $B(k-1)$, which has an asymptotic form \cite{oeis_bell, wilf_bell}
	
	\begin{equation}
		\label{eqn:bell}
		B(k-1) = \sum_{m=1}^{k-1} m! \ S(k-1, m) \stackrel{k \gg 1}{\approx} \frac{(k-1)!}{2(\ln{2})^k}.
	\end{equation}
	
	\noindent So we obtain to lowest order in $Q$,
	\begin{equation}
		\label{eqn:star_d_ex1}
		D(k) \approx Q^{k-1} B(k-1).
	\end{equation}
	
	To show the existence of a minimum, we explicitly estimate $k_{star}^*$ using the asymptotic form of the ordered Bell number (Eq.~(\ref{eqn:bell})). The use of the asymptotic form is justified by the observation that $k_{star}^*$ is large for small $Q$ (Fig.~\ref{fig:star}), so for $Q \ll 1$ we have $k_{star}^* \gg 1$. Substituting Eq.~(\ref{eqn:bell}) into Eq.~(\ref{eqn:star_d_ex1}), we obtain
	
	\begin{equation}
		\label{eqn:star_d_expanded}
		D(k) \approx Q^{k-1} \frac{(k-1)!}{2 (\ln{2})^{k-1}}.
	\end{equation}
	
	\noindent Minimizing this function with respect to $k$ gives us an estimate for the optimum point $k_{star}^*$. The derivation is shown in detail in the Appendix, and the result is 

	\begin{equation}
		\label{eqn:k_star}
		k_{star}^* \approx \frac{\ln{2}}{Q}.
	\end{equation}
	
	\begin{figure}[t]
		\includegraphics[width=\columnwidth]{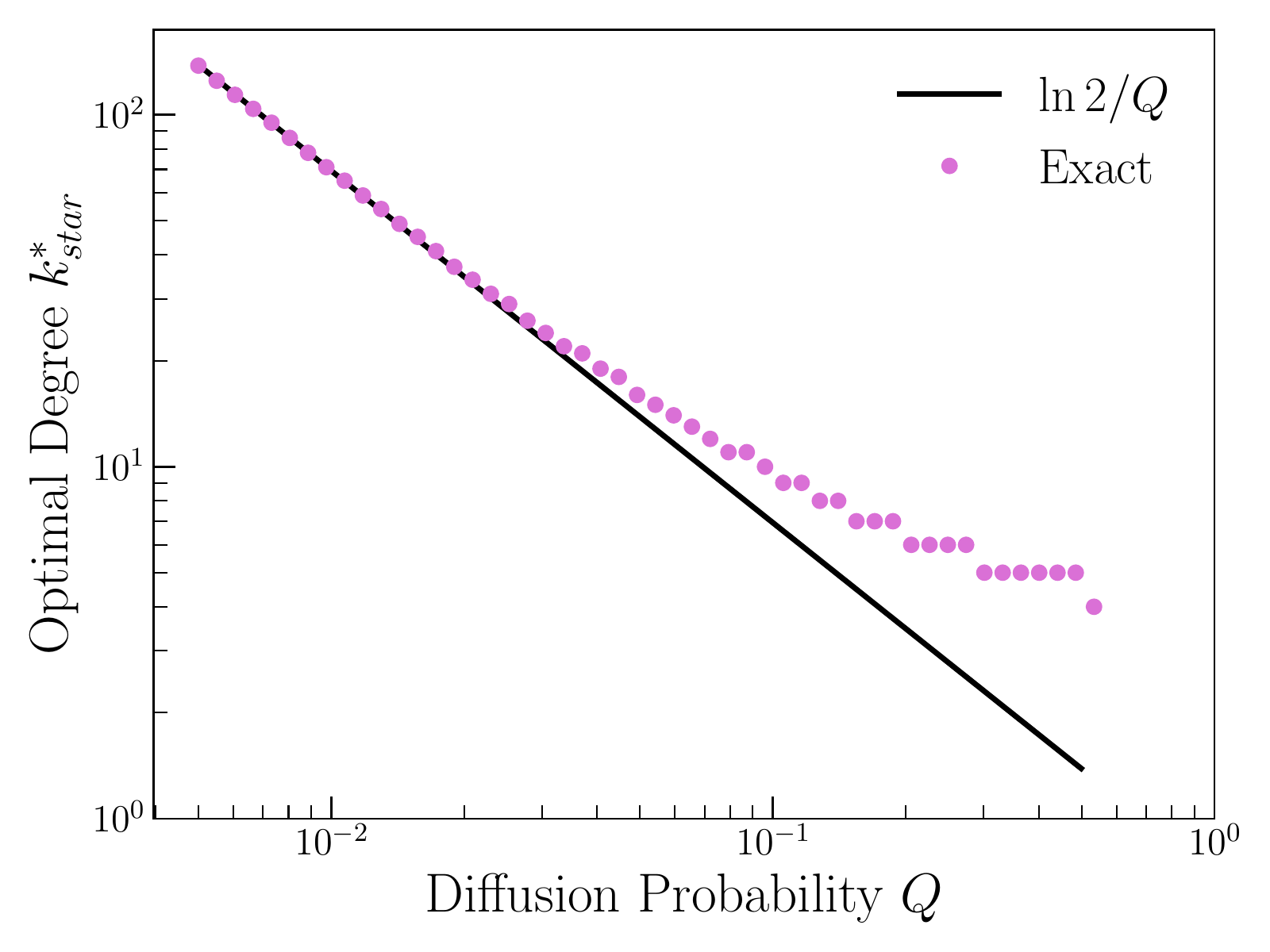}
		\centering	
		\caption{\label{fig:kstar} The star graph optimal degree $k_{star}^*$ as estimated by Eq.~(\ref{eqn:k_star}) (solid line) compared with its exact value obtained from numerically solving Eqs.~(\ref{eqn:star_trans})-(\ref{eqn:star_recur}). Our estimation shows good agreement with direct numerical solution for $Q<0.1$.}
	\end{figure}	
	
	Fig.~\ref{fig:kstar} compares our estimate using Eq.~(\ref{eqn:k_star}) to the exact $k_{star}^*$ obtained from numerically solving the Markov chain Eqs.~(\ref{eqn:star_trans})-(\ref{eqn:star_recur}) for a range of $Q$. The results suggest that Eq.~(\ref{eqn:k_star}) does provide a good estimate of $k^*$ for $Q < 0.1$. This combinatorial construction also offers an intuitive explanation for the existence of $k_{star}^*$: The form of Eq.~(\ref{eqn:star_d_ex1}) suggest that as the number of link $k$ increases, there is a competition between the diminishing probability for infecting more links and the growing number of ``dead-end patterns'' that will kill the node. When $k<k_{star}^*$, the suppression factor $Q^{k-1}$ dominates, and adding more links would decrease the dead probability. When $k>k_{star}^*$, the combinatorial growth of $B(k-1)$ eventually catches up, and $Q^{k-1}$ can no longer suppress all the new dead-end patterns created by adding new links. At $k^*$, the two effects balance, and one achieves the lowest possible dead probability. Finally, when one is far above $k_{star}^*$, the combinatorial growth is so overwhelming that there are simply too many possible dead-end patterns that a node will possibly encounter, and thus $D(k \to \infty) \to 1$ for all $Q>0$.

	Although in this work we define a dead node to be one that lost all its links, it is also possible to soften this definition and consider instead a death threshold $z$, such that a node is considered dead whenever the fraction of its inactive links exceeds $z$. The results with this definition of death are however similar to the results we presented using the strict definition (which corresponds to $z=1$). In particular, our combinatorial argument for the existence of $k^*$ is still valid, and simulations show the persisted existence of $k^*$ for $z < 1$.
	
	Having understood the features of our model on simple star graphs, we now turn to the analysis of our model on more general networks.
	
\section{\label{sec:nwk}Dynamics on General Networks}
	\subsection{\label{sec:nwk_1}Mean-field Theory}
	A formal mathematical treatment of our model on general networks is considerably more difficult because of the many degrees of freedom involved. We here formulate a mean-field theory as an approximation of our model and show that it is sufficient to capture the basic dynamical features, including the appearance of the minimum point $k^*$.

	Following our analytic treatment of the star network, we assign to each node $i$ a state variable $x_i = (n_i, s_i)$, with $n_i$ and $s_i$ being respectively the number of active and susceptible links of a node $i$. The model dynamics is fully parametrized by the set $\{ x_i \} = \{x_1, \dots, x_N \}$, where $N$ is the total number of nodes in the network. The full dynamics can be written in the form of a many-body Markov process in the state distribution $P(\{x_i \}, t)$ via a many-body transition probability $T$ as
	
	\begin{equation}
		\label{eqn:nb_master}
		P\left(\{x_i \}, t+1\right) = \sum_{\{x_i' \}} T\left(\{x_i \} \mid \{x_i' \} \right)P\left(\{x_i' \}, t\right).
	\end{equation}
	
	\noindent To obtain an equation for the single-node state distribution $P(x_a, t)$ for a particular node $a$, we sum Eq.~\ref{eqn:nb_master} over all irrelevant state variables $x_i$ with $i\neq a$, yielding
	  
	\begin{equation}
		\label{eqn:nb_master_summed}
		P \left(x_a, t+1 \right) = \sum_{\{ x_i' \}} \left( \sum_{\{x_{i \neq a}\}} T\left(\{ x_i \} \mid \{ x_i' \}\right) \right) P \left(\{ x_i' \}, t\right).
	\end{equation}
	
	\noindent The factor in the parenthesis is the net transition probability for a single-node transition $x_a' \to x_a$, regardless of what the rest of the state variables will transit to. Since a link can be infected only by a neighboring link, one expects physically that transitions of node $a$ can only depend on the states of node $a$ itself and of its nearest neighbors. Conversely, for nodes not in the nearest neighborhood of $a$, since their links are not directly connecting to links of $a$, these nodes cannot affect the transition of $a$ in one time step. Thus, denoting the set of neighbor states around node $a$ to be $\{x_{\mathcal{N}(a)} \}$, we write
	
	\begin{equation}
		\label{eqn:nb_trans_single}
		 \sum_{\{x_{i \neq a}\}} T\left(\{ x_i \} \mid \{ x_i' \}\right) = T\left(x_a \mid x_a', \{ x_{\mathcal{N}(a)}' \} \right).
	\end{equation}
	
	\noindent The validity of Eq.~(\ref{eqn:nb_trans_single}) will be confirmed as we construct the explicit form of $T\left(x_a \mid x_a', \{ x_{\mathcal{N}(a)}' \} \right)$ later on. After substituting Eq.~(\ref{eqn:nb_trans_single}) into Eq.~(\ref{eqn:nb_master}), the $\{x_i'\}$ sum can now be partially performed to give
	
	\begin{align}
		\label{eqn:mb_single}
		P\left(x_a, t+1 \right) = \sum_{x'_a} & \sum_{\{x_{\mathcal{N}(a)}' \}} T\left(x_a \mid x'_a, \{x'_{\mathcal{N}(a)} \} \right) \nonumber \\
		 & \times P\left(x'_a, \{x'_{\mathcal{N}(a)} \}, t\right).
	\end{align}

	Eq.~(\ref{eqn:mb_single}) is an exact equation relating the single-body distribution $P(x_a, t)$ to the neighborhood distribution $P(x_i, \{x_{\mathcal{N}(a)}\}, t)$. In principle, one could repeat the same procedure and obtain an equation relating the nearest neighborhood distribution to the next-nearest neighborhood distribution, then to the next-next-nearest neighborhood and so on. This is essentially a statistical closure problem similar to solving the BBGKY hierarchy in classical kinetic theory \cite{liboff_kinetic}, and doing so is equivalent to solving the full many-body problem which is impractical.
	
	To truncate this hierarchy, we now formulate a mean-field approximation to systematically neglect correlations so that a closed system of equations can be obtained. Following the degree-based mean-field approach of Pastor-Satorras and Vespignani \cite{pastor-satorras_internet, pastor-satorras_epidemic, boguna_mf, barrat_dynamics}, we first divide the nodes into classes based on their degree $k$ and neglect all network structures except the degree distribution and degree-degree correlation. Within this approximation, all nodes in the same $k$-class are assumed to be statistically identical. We therefore replace the original state distributions $P(x_a, t)$ by a set of $k$-dependent state distributions $P_k(x, t)$, one for each $k$-class. Furthermore, we assume connections among nodes depend solely on the degree-degree correlation with no further structure, so we may characterize connections by the conditional probability $C(k'|k)$ that a link starting from a $k$-degree node is connected to a $k'$-degree node \cite{barrat_dynamics, boguna_mf}. Applying these approximations, and using Greek indices to label neighboring nodes (so that a $k$-degree node has neighbors indexed by $\alpha = 1, \dots, k$), we can rewrite the single-body evolution equation Eq.~(\ref{eqn:mb_single}) in a more suggestive form
	
	\begin{equation}
		\label{eqn:mf_master}
		P_k \left(x, t+1 \right) = \sum_{x'} \bar{T}_k\left(x, t+1 \mid x', t \right) P_k \left(x', t \right),
	\end{equation}
	
	\noindent with $\bar{T}_k\left(x, t+1 \mid x', t \right)$ being an effective single-body transition probability given by
	
	\begin{align}
		\label{eqn:mf_Tbar}
		\bar{T}_k(x, t+1 \mid x', t ) = \sum_{\{x_\alpha' \}}& T\left(x \mid x', \{x_\alpha'\} \right) \nonumber \\
		 & \times P_k\left( \{x_\alpha'\}, t \mid x', t \right).
	\end{align}

	To complete the mean-field approach, we need to supply a suitable approximation for the neighborhood distribution $P_k\left( \{x_\alpha'\}, t \mid x', t \right)$ in terms of the single-body distribution $P_k(x, t)$ and evaluate the sum in Eq.~(\ref{eqn:mf_Tbar}), so that a closed set of equations for $P_k(x, t)$ can be obtained.	

	Ignoring dynamical correlations amount to factorizing the neighborhood distribution function into products of single-body distributions. This invites the mean-field ansatz
	\begin{equation}
		\label{eqn:mf_simple}
		P_k\left(\{x_\alpha \}, t \mid x, t\right) = \prod_{\alpha=1}^k \sum_{l_\alpha} C(l_\alpha \mid k) P_{l_\alpha}(x_\alpha, t),
	\end{equation}
	\noindent where $C(l_\alpha | k)$ is the conditional connection probability introduced above and normalizes as $\sum_{l_\alpha=1}^{k_{max}} C(l_\alpha | k) = 1$, with $k_{max}$ being the maximum degree of the network. This ansatz will however assign non-zero probability to physically impossible configurations. For example, if the central node has state $x = (n=0, s=k)$, then it is impossible to have a configuration where any of its neighbors has $x_\alpha = (n_\alpha, s_\alpha = 0)$, since the central node can only be connected through susceptible links. These physical constraints are not respected by Eq.~(\ref{eqn:mf_simple}). To derive an appropriate form of Eq.~(\ref{eqn:mf_simple}) that respects such physical constraints, we introduce the following ordering convention for the neighbor indices $\alpha$: We choose to arrange the neighbors in such a way that the first $s$ of them ($\alpha = 1, \dots, s$) are the ones connected to the central nodes via a susceptible link, the next $n$ of them ($\alpha = s+1, \dots, s+n$) are connected via an active link, and the remaining $k-n-s$ ($\alpha = s+n+1, \dots, k$) are connected via an inactive link. This convention is schematically summarized as follows
	\begin{equation}
		\label{eqn:ordering_convent}
		\{\alpha \} \equiv \{ \underbrace{1, \dots , s}_{\substack{\text{$s$  susceptible} \\ \text{connections}}}, \ \underbrace{s+1, \dots , s+n}_{\substack{\text{$n$ active} \\ \text{connections}}}, \  \underbrace{s+n+1, \dots , k}_{\substack{\text{$(k-n-s)$ inactive} \\ \text{connections}}} \}.
	\end{equation}
	
	To impose the aforementioned physical constraints, we make the observation that without further information, one expects the probability of connecting to a node via a susceptible (active, inactive) link to be simply proportional to the node’s number of susceptible (active, inactive) links. More precisely, one expects the probability $\Gamma_0(x, t \mid k)$ ($\Gamma_1(x, t \mid k)$, $\Gamma_2(x, t \mid k)$) of connecting from a $k$-degree node to a node with state $x = (s, n)$ via a susceptible (active, inactive) link to be
	
	\begin{subequations}
	\label{eqn:mf_gamma}
	\begin{align}
		\Gamma_0(x, t \mid k) &= \sum_l C(l \mid k) \ \frac{s P_{l}(x, t)}{\sum_{x'} s' P_l(x', t)}, \\
		\Gamma_1(x, t \mid k) &= \sum_l C(l \mid k) \ \frac{n P_{l}(x, t)}{\sum_{x'} n' P_l(x', t)}, \\
		\Gamma_2(x, t \mid k) &= \sum_l C(l \mid k) \ \frac{r P_{l}(x, t)}{\sum_{x'} r' P_l(x', t)},
	\end{align}
	\end{subequations}
	
	\noindent where $r = (l - s - n)$ is the number of inactive links. Note that 

	\begin{align}
		\label{eqn:mf_gamma_norm}
		\sum_x \Gamma_0(x, t \mid k) &= \sum_x \sum_l C(l \mid k) \ \frac{s P_{l}(x, t)}{\sum_{x'} s' P_l(x', t)} \nonumber \\
			&= \sum_l C(l \mid k) \left( \sum_x \ \frac{s P_{l}(x, t)}{\sum_{x'} s' P_l(x', t)} \right) \nonumber \\
			&= \sum_l C(l \mid k) = 1,
	\end{align}
	
	\noindent where in the second equality, we have interchanged the order of summation. This can be done since both $l$ and $x$ are bounded by the maximum degree $k_{max}$ in the network, and we are summing over finite series and thus can rearrange the terms and interchange the summation order. Similar conclusions also apply to $\Gamma_1, \Gamma_2$, and hence $\Gamma_0, \Gamma_1, \Gamma_2$ are properly normalized probabilities with respect to the state $x$ of the node to be connected. Using Eq.~(\ref{eqn:mf_gamma}) and the ordering convention in Eq.~(\ref{eqn:ordering_convent}), we now propose the physically consistent ansatz 
	
	\begin{align}
		\label{eqn:mf_constrained}
	P_k(\{x_\alpha \}, t \mid x, t ) =  &\prod_{\alpha = 1}^{s} \Gamma_0(x_\alpha, t \mid k) \prod_{\beta = s+1}^{s+n} \Gamma_1(x_\beta, t \mid k) \nonumber \\*
		 &\times \prod_{\gamma = s+n+1}^{k} \Gamma_2(x_\gamma, t \mid k).
	\end{align}

	In order to use this mean-field ansatz to evaluate the effective single-body transition probability $\bar{T}_k$, Eq.~(\ref{eqn:mf_Tbar}), we still need the explicit form of the exact transition probability $T(x \mid x', \{x_\alpha' \})$. This can be done by modifying the star graph transition probability $T_{star}$ to include effects from nearest neighbors. We start by rewriting $T_{star}$ (Eq.~(\ref{eqn:star_trans})) as 
	
	\begin{align}
		\label{eqn:star_trans_rewrite}
		T_{star} \big(x \mid x' \big) = \frac{\delta_{s, s'-n}}{n! \ s!} &\sum_{\sigma} \prod_{\mu=1}^{n} \big(1 - (1-Q)^{n'} \big) \nonumber \\* 
		&\times \prod_{\nu=n+1}^{n+s} \big(1-Q \big)^{n'},
	\end{align}

	\noindent where the sum is over all permutations $\sigma$ on the set of integers $\{1, \dots, s' \}$. Since nothing in this expression depends explicitly on $\sigma$, the $\sigma$-sum trivially produces a factor of $s'!$, which gives back the original expression (Eq.~(\ref{eqn:star_trans})). We may reinterpret the product indices $\mu$ and $\nu$ to be indices labeling the $s'$ original susceptible links, so that the products pick the first $n$ links to infect and leave the remaining $s$ links untouched. Since we only care about the total number of links chosen to get infected, we therefore sum over all configurations or permutations of $\sigma$ where the links are being placed to obtain the total probability of infecting $n$ links. The $n!(s'-n)!$ is a combinatorial factor to avoid overcounting. For the star graph, since the nearest neighbors have no structure, all the links are identical, and the sum simply produces a binomial coefficient. 
	
	To write down the many-body transition probability, one should note that as one includes the full network structure, the links are no longer identical. This is because the neighboring nodes can have links connect to other nodes and thus each neighboring node has a different state. Hence the infection probability of a link is now affected by both the states of the central node and of the neighboring nodes, which gives
	
	\begin{align}
		\label{eqn:nb_trans_explicit}
		T \big(x \mid x', \{x'_\alpha \} \big) = \frac{\delta_{s, s'-n}}{n!\ s!} &\sum_{\sigma} \prod_{\mu=1}^{n} \big(1 - (1-Q)^{n' + n'_{\sigma(\mu)}} \big) \nonumber \\*
		&\times \prod_{\nu=n+1}^{n+s} \big(1-Q \big)^{n' + n'_{\sigma(\nu)}}, 
	\end{align}

	\noindent where $n'_{\sigma(\mu)}, n'_{\sigma(\nu)}$ denote the number of active links of the neighboring nodes. Note that nodes farther away from nearest neighbors can have no effects since they are not directly connected to links of the central node. Therefore, condition Eq.~(\ref{eqn:nb_trans_single}) is naturally implied, which indicates that our formulation is self-consistent.
	
	With Eq.~\ref{eqn:nb_trans_explicit}, we can now evaluate $\bar{T}_k$ (Eq.~(\ref{eqn:mf_Tbar})) using the mean-field ansatz of Eq.~(\ref{eqn:mf_constrained}). Substituting Eq.~(\ref{eqn:nb_trans_explicit}) and Eq.~(\ref{eqn:mf_constrained}) into Eq.~(\ref{eqn:mf_Tbar}) gives 

	\begin{align}
		\label{eqn:Tbar_substituted1}
		\bar{T}_k &= \frac{\delta_{s, s'-n}}{n! s!} \sum_{x'_1} \cdots \sum_{x'_k} \sum_\sigma \nonumber \\*
		&\times \prod_{\mu=1}^{n} \left[ \left(1 - (1-Q)^{n'+n'_{\sigma(\mu)}} \right) \Gamma_0(x'_{\sigma(\mu)}, t \mid k) \right] \nonumber \\*
		&\times \prod_{\nu=n+1}^{n+s} \left[ \left((1-Q)^{n'+n'_{\sigma(\nu)}}\right) \Gamma_0(x'_{\sigma(\nu)}, t \mid k) \right]\nonumber \\*
		&\times \prod_{\beta=s'+1}^{s'+n'} \Gamma_1(x'_\beta, t \mid k) \prod_{\gamma=s'+n'+1}^{k} \Gamma_2(x'_\gamma, t \mid k),
	\end{align}

	\noindent where $\sum_\sigma$ denotes the sum over all permutations on the indices $\{1, \dots, s'=n+s \}$. The $\Gamma_0$-product in Eq.~(\ref{eqn:mf_constrained}) is now combined with the products in Eq.~(\ref{eqn:nb_trans_explicit}), since both of them are products over the neighbor indices $\{1, \dots, s'=n+s \}$.
	
	We now interchange the order of summations and products to perform the sums over $x'_1 \dots x'_k$, again justified by the fact that Eq.~(\ref{eqn:Tbar_substituted1}) is a finite series. Each factor in the last two products sums to $1$ by the normalization condition Eq.~(\ref{eqn:mf_gamma_norm}), and we are left with

	\begin{align}
		\label{eqn:Tbar_substituted2}
		\bar{T}_k &= \frac{\delta_{s, s'-n}}{n! s!} \sum_\sigma \nonumber \\*
		&\times \prod_{\mu=1}^{n} \left[\sum_{x'_{\sigma(\mu)}} \left(1 - (1-Q)^{n'+n'_{\sigma(\mu)}} \right) \Gamma_0(x'_{\sigma(\mu)}, t \mid k) \right] \nonumber \\*
		&\times \prod_{\nu=n+1}^{n+s} \left[\sum_{x'_{\sigma(\nu)}} \left((1-Q)^{n'+n'_{\sigma(\nu)}}\right) \Gamma_0(x'_{\sigma(\nu)}, t \mid k) \right].
	\end{align}
	
	\noindent Eq.~(\ref{eqn:Tbar_substituted2}) does not contain $\Gamma_1$ and $\Gamma_2$ since infection only occurs on susceptible links, and the status of already infected links (which are described by $\Gamma_1, \Gamma_2$) are therefore irrelevant. Furthermore, each factor in the $\mu$- and $\nu$-products leads to the same sum, regardless of the presence of $\sigma$. Therefore, if we define the quantity

	\begin{align}
		\label{eqn:mf_qeff}
		Q_{eff}(x', t; k) \equiv \sum_{x_\mu'} \left(1 - (1-Q)^{n'+n_\mu'} \right) \Gamma_0(x'_\mu, t \mid k),
	\end{align}
	
	\noindent we obtain from the $\mu$-products $n$ factors of $Q_{eff}$ and from the $\nu$-products $s$ factors of $(1 - Q_{eff})$. Thus, we have
	
	\begin{equation}
		\bar{T}_k(x, t+1 \mid x', t) = \frac{\delta_{s, s'-n}}{n! s!} \sum_\sigma Q_{eff}^{n} \left( 1 - Q_{eff} \right)^{s}.
	\end{equation}
	
	\noindent Finally, since there is no dependence on $\sigma$ explicitly, summing over $\sigma$ simply generate a factor of $s'! = (n+s)!$. Combining this with $1/(n!\ s!)$ and rewriting $s = s'-n$ leads to our final result
	
	\begin{equation}
		\label{eqn:mf_trans}
		\bar{T}_k(x, t+1 \mid x', t) = \delta_{s, s'-n} \binom{s'}{n} Q_{eff}^n (1 - Q_{eff})^{s'-n},
	\end{equation}
	
	\noindent with the binomial parameter $Q_{eff}$, which acts as an effective infection probability, given by Eq.~(\ref{eqn:mf_qeff}).

	Eqs.~(\ref{eqn:mf_qeff}) and (\ref{eqn:mf_trans}), together with Eq.~(\ref{eqn:mf_master}) constitute a closed set of equations that can be numerically solved for the state distribution functions $P_k(x, t)$. For the single seed link scenario we are considering, these equations should be solved with the initial condition $P_k(x, t=0) = \frac{k}{E} \delta_{n, 1} \delta_{s, k-1} + (1 - \frac{k}{E}) \delta_{n, 0} \delta_{s, k}$, where $E$ is the total number of links in the network.

	\begin{figure*}[t]
		\includegraphics[width=\textwidth]{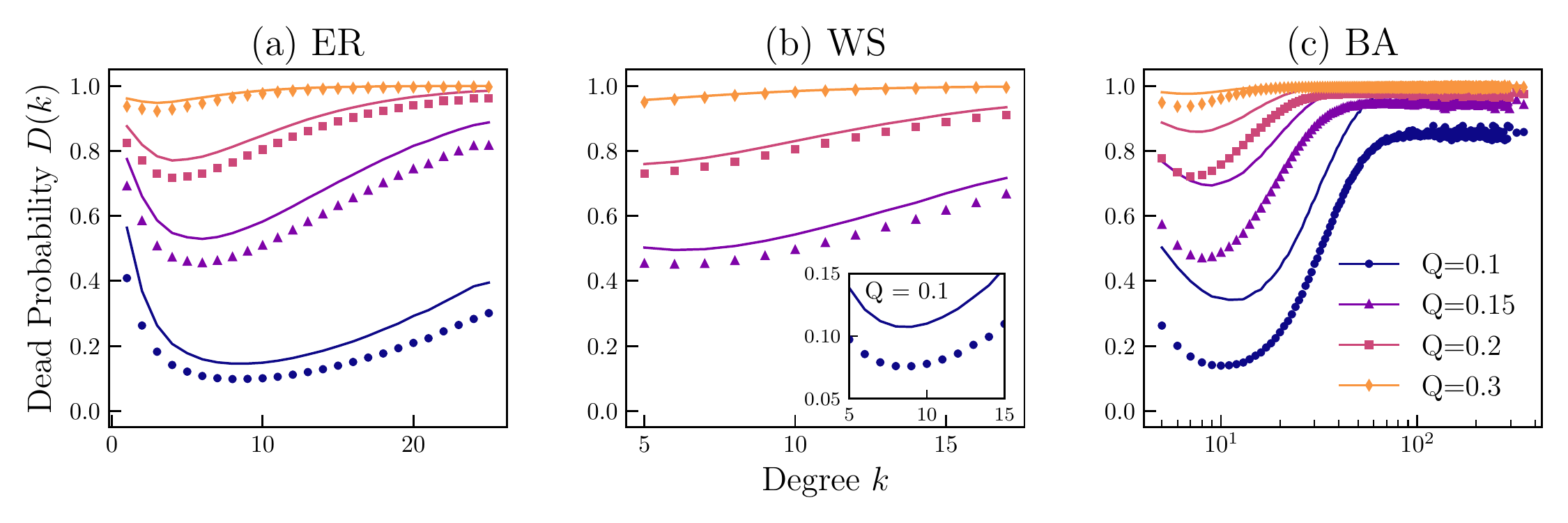}
		\centering	
		\caption{\label{fig:nwk_models} Direct simulation of our link cascade model (markers), compared with mean-field theory (solid lines) on (a) Erdős–Rényi (ER) random networks, (b) Watts-Strogatz (WS) small world networks, and (c) Barabási–Albert (BA) scale-free networks, for four different values of diffusion probability $Q$. The inset in (b) is the curve for $Q=0.1$ on WS network, singled out to better display the minimum point of the curve. All generated networks have $N=5000$ nodes and average degree $\langle k \rangle = 10$. The WS networks have a rewiring probability $p_{WS}=0.3$. All three network models show qualitatively similar behavior as the star graph. In particular, minimal points $k^*$ are observed in the ER and BA networks for all $Q$. The mean-field theory is in close agreement with simulation for ER and BA networks, although the discrepancy is somewhat larger for BA networks.}
	\end{figure*}

	\subsection{\label{sec:nwk_2}Optimum Degree in General Networks}
	
	Based on the mean-field theory, we now argue that a minimum point in dead probability $D(k)$, similar to the star graph should be expected for general networks. Define a time-dependent average 

	\begin{equation}
		\label{eqn:mf_navg}
		\langle f(x) \rangle_{k, t} = \sum_{x} f(x) \ \Gamma_0(x, t \mid k), 
	\end{equation}

	\noindent so that $\langle f(x) \rangle_{k, t}$ is the average value of a state function $f(x)$ of the $k$-degree node as seen by its neighbors that are connected by susceptible links at time $t$. With this definition, Eq.~(\ref{eqn:mf_qeff}) can be rewritten as
	
	\begin{equation}
		\label{eqn:mf_Qeff_navg}
		Q_{eff}(x', t;k) = 1 - \left(1 - Q \right)^{n'} \langle\left(1 - Q \right)^n \rangle_{k, t}.
	\end{equation}

	Again considering the $Q \ll 1$ limit, Eq.~(\ref{eqn:mf_Qeff_navg}) can be expanded to lowest order in $Q$ as
	
	\begin{align}
		\label{eqn:mf_Qeff_expanded}
		Q_{eff}(x', t;k) &= 1 - \left(1 - n'Q + \dots \right) \langle 1 - nQ + \dots \rangle_{k, t} \nonumber \\*
		&= n'Q + \langle n\rangle_{k, t}Q + \mathcal{O}(Q^2) \nonumber \\&
		\approx n'\left(1 + \frac{\langle n\rangle_{k, t}}{n'} \right)Q .
	\end{align}
	
	\noindent Eq.~\ref{eqn:mf_trans} then becomes, in the $Q \ll 1$ limit,
	
	\begin{equation}
		\label{eqn:mf_trans_expanded}
		\bar{T}_k(x, t+1 \mid x', t) \approx \delta_{s, s'-n} \binom{s'}{n} n'^n \left(1 + \frac{\langle n\rangle_{k, t}}{n'}\right)^n Q^n.
	\end{equation}
	
	\noindent This has essentially the same form as the star graph case in Eq.~(\ref{eqn:star_trans_expanded}), except that $Q$ has acquired a time-dependent enhancement factor $(1 + \langle n \rangle_{k, t} / n')$. Without further assumptions, one cannot obtain an analytic form of this enhancement factor, and thus will affect the estimation of $D(k)$. However, our earlier combinatorial argument for the existence of $k^*$ still applies -- there will be a competition between the exponential suppression from factors of $Q$ versus the combinatorial growth of infection patterns and the balance between the two effects will yield a minimum dead probability. Therefore, in the mean-field limit, the appearance of $k^*$ is a pure combinatorial effect that applies generally to any network. The exact location of $k^*$ will depend on the precise details of the network through the factor $(1 + \langle n \rangle_{k, t} / n')$. Depending on the network topology, $k^*$ may take values less than the minimum degree $k_{min}$ in the network. In this case $D(k)$ will behave as an increasing function over the whole degree spectrum and we simply have $k^* = k_{min}$.

	\subsection{\label{sec:nwk_3}Numerical Simulations}

	\begin{figure*}[t]
		\includegraphics[width=\textwidth]{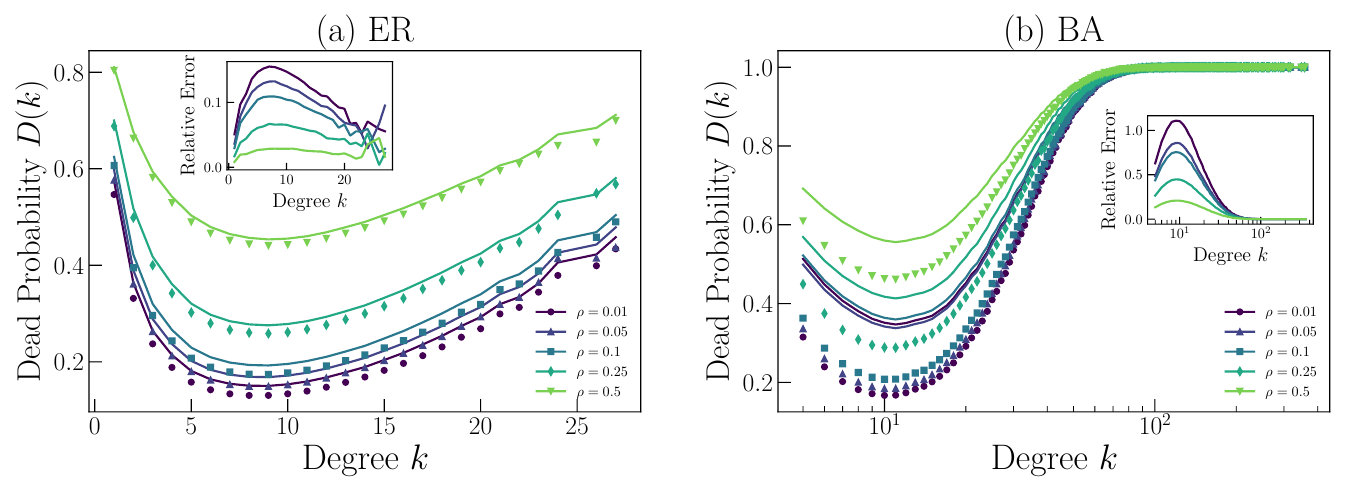}	
		\centering
		\caption{\label{fig:multiple} Comparison of direct simulation (markers) with mean-field theory (solid lines) for multiple-seed link cascade with seed-link fraction $\rho = 0.01, 0.05, 0.1, 0.25, 0.5$ on (a) ER networks and (b) BA networks. All generated networks have $N=5000$ nodes and average degree $\langle k \rangle = 10$, and the diffusion probability is set to be $Q=0.1$. The insets show the relative error of the mean-field approximation with respect to simulation. The behavior of $D(k)$ is qualitatively unaltered by the value of $\rho$. As $\rho$ increases, the difference between mean-field theory and simulation become smaller.}
	\end{figure*}

	We now examine our link independent cascade model on several well-studied network models by directly simulating the dynamical rules (Sec.~\ref{sec:model}) on the networks and compare the simulation results with the mean-field theory that we have formulated. Fig.~\ref{fig:nwk_models} shows the results, on (a) Erdős–Rényi (ER) random networks, (b) Watts-Strogatz (WS) small world networks, and (c) Barabási–Albert (BA) scale-free networks. $20$ different network realizations are generated for each network model. All networks generated for this study are chosen to have $N=5000$ nodes and an average degree of $\langle k \rangle = 10$. For the WS networks, the rewiring probability is chosen to be $p_{WS}=0.3$ in order to obtain a moderately broad degree distribution while still maintaining its characteristic high clustering property \cite{barrat_WS}. The dead probability $D(k)$ is computed statistically from averaging over $1000$ simulation runs per network realization, giving a total of $20000$ simulation runs per network model, for each of $Q=0.1, 0.15, 0.2, 0.3$. We also plot $D(k)$ obtained from mean-field theory (solid curve in Fig.~\ref{fig:nwk_models}) to compare with direct simulation results (markers in Fig.~\ref{fig:nwk_models}). The mean-field prediction is in close agreement with simulation results for the ER and WS networks, while the discrepancy is somewhat larger for the BA network. To obtain $C(l \mid k)$ for finite ER, WS, and BA networks, we have directly estimated $C(l \mid k)$ from the generated network realizations, and as a result of this estimation, the mean-field curves are not completely smooth. The mean-field theory tends to overestimate the efficiency of the spreading hence the value of $D(k)$. This is because our mean-field approximation neglects dynamical correlations stemmed from the localized nature of the cascade process and introduces spurious long-range interactions that are not present in the original system. In the original definition of the model, an infected link can only infect its nearest neighboring links, and thus a link being a distance $d$ away from the seed link will not be infected until $t=d$. In our mean-field theory, such distance-induced correlations have not been explicitly treated. The influence of infected links is instead received by all links in the network equally irrespective of their location, which effectively allows long-range interaction to take place. These spurious long-range interactions would cause links to become more easily and frequently infected, and further allow a ``back-propagation’’ of infection to re-enter regions where the infection has already passed through and left behind surviving nodes. Consequently, the dead probability is overestimated in our mean-field theory. This overestimation is more severe for the BA network, possibly because of the presence of more high degree nodes, which tend to have a large number of infected links and amplify the effect from the mean-field approximation outlined here. The construction of a more accurate mean-field theory for scale-free networks will be a topic for future work.

	The curves in Fig.~\ref{fig:nwk_models} share most of the basic features of the star graph results in Fig.~\ref{fig:star} and matches our expectation from the mean-field theory in Sec.~\ref{sec:nwk_2}. All three networks show an increasing dead probability in the high degree end, and in particular, both the ER and BA networks have an optimum degree $k^*$ for all $Q$. For the WS network, since its degree distribution is concentrated around $\langle k \rangle$, the expected optimal degree may have fallen out of the available range of $k$ in some cases, and thus one is unable to observe a $k^*$. Furthermore, due to its narrow degree distribution when compared to an ER network with the same $\langle k \rangle$, the initial seed link has a much smaller probability to land on a high degree node, which may cause an overall weaker spreading in the WS network, as seen in its lower dead probability in the figure. In contrast, the BA network shows the strongest degree dependence due to its broad degree distribution. Based on these results, we believe that scale-free networks are most susceptible to link cascade damage in real-world applications.

	\subsection{\label{sec:nwk_4}Multiple-seed Cascade}
	We can now extend the treatment to the multiple-seed case when we have more than one seed to trigger a cascade across the network. The multiple-seed case should give a result closer to that obtained from the mean-field approximation, since seeds far apart resemble the long-range interactions in the mean-field approximation and induce a ``back-propagation’’ effect on the nodes. It is conceivable that the multiple-seed case is more relevant to social networks with strong community structures, since in such networks link cascades are more likely to occur within individual communities. In this case, the result will roughly be the sum of link cascades in different communities.

	Our formulation of mean-field theory is still valid for the multiple-seed case, except that the mean-field equations need to be solved with an appropriate multiple-seed initial condition. We assume that the seed links are selected independently with uniform selection probability, and we denote $e = \rho E$ to be the number of seed links. There are different forms one can use for the multiple-seed initial state but we here generalize our single-seed initial state to obtain the multiple-seed initial state distribution $P_k(x, t=0)$ by computing the probability that $n$ of the $e$ seed links are selected to be any of the $k$ links that a $k$-degree node possesses. This is combinatorially equivalent to the probability of drawing without replacement $n$ red balls in $e$ trials from an urn of $E$ balls, of which $k$ of them are red, and this probability is given by the hypergeometric distribution \cite{feller_probability}. Noting further that after $n$ links are selected to be seeds, there will remain $s = k-n$ susceptible links, we write the multiple-seed initial condition as

	\begin{equation}
		\label{eqn:ms_initial}
		P_k(x, t=0) = \frac{\binom{k}{n} \binom{E-k}{e-n}}{\binom{E}{e}} \delta_{s, {k-n}}.
	\end{equation}

	\noindent For the case of $e=1$, Eq.~(\ref{eqn:ms_initial}) reduces back to the single-seed initial condition.

	Fig.~\ref{fig:multiple} shows the comparison of mean-field theory with direction simulation for multiple-seed cascade at different values of seed link fraction $\rho$. For comparison, we again generated $20$ realizations of ER and BA networks with $N=5000$ and $\langle k \rangle = 10$ and performed $1000$ simulation runs per network realization. In the simulation, we fix the value of diffusion probability to be $Q=0.1$. The insets plot the relative error of mean-field theory with respect to simulation. The results show that choosing different values of $\rho$ does not qualitatively change the behavior of $D(k)$. As $\rho$ increases, the difference between mean-field theory and simulation become smaller, meaning that the multiple-seed case is closer to the mean-field limit as we have claimed.

	\begin{table*}[t]
		\caption{\label{table:topology}Summary of network topology, including number of nodes $N$, number of links $E$, network density $2E / (N(N-1))$, average degree $\langle k \rangle$, degree standard deviation $\sqrt{\langle (k-\langle k\rangle)^2 \rangle}$, and average clustering coefficient $\langle c \rangle$, for the Slashdot network. For more details see Ref.~\onlinecite{leskovec_slashdot}.} 
		\begin{ruledtabular}
		\begin{tabular}{cccccc}
		
		$N$ & $E$ & $\frac{2E}{N(N-1)}$ & $\langle k \rangle$ & $\sqrt{\langle (k-\langle k\rangle)^2 \rangle}$ & $\langle c \rangle$ \\
		\hline
		$77350$ & $468554$ & $1.566\times 10^{-4}$ & $12.115$ & $40.395$ & $0.055$ \\
		\end{tabular}
		\end{ruledtabular}
	\end{table*}

	\begin{figure*}[t]
		\includegraphics[width=\textwidth]{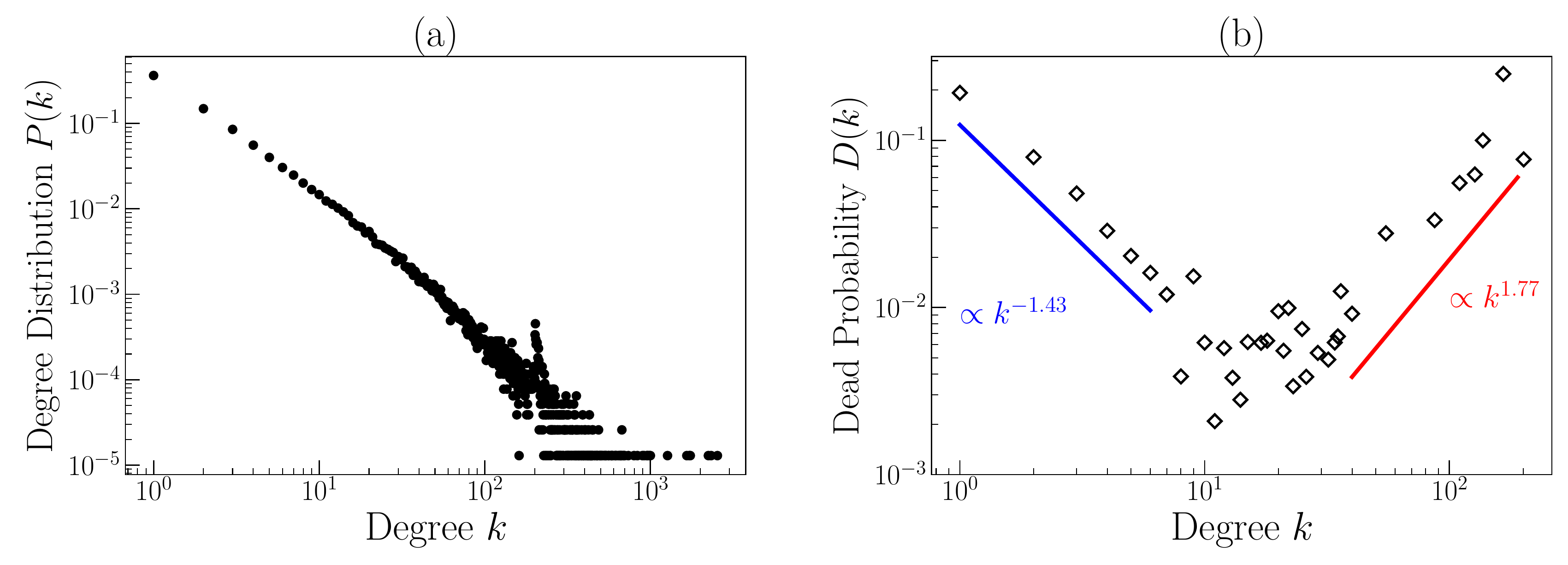}					
		\centering
		\caption{\label{fig:data} (a) Degree distribution $P(k)$ of the Slashdot network, which has a power law form. (b) Dead probability $D(k)$ in the Slashdot network measured by counting the number of nodes with all negative links. A minimum point $k^*$ can be observed and $D(k)$ can be approximated by power law fits around $k^*$.}
	\end{figure*}
\section{\label{sec:app}Application: Friend-Enemy Networks}

	In this section, we present an application of our link cascade model to the problem of friend-enemy network formation dynamics. We consider a social network $\mathcal{G}$, with each of its links representing a possible social interaction or connection between a pair of individuals. We may then indicate the social relationships between individuals by assigning to each link a numerical sign ($+1$ or $-1$), where a positive sign represents friendship or trust and a negative sign represents hostility or distrust. Given such a setup, one natural question to ask is whether one can model, or even predict the distribution of these hostile negative links. More specifically, one may ask what is the probability for one to become a ``persona non grata'', that is, an individual who has no friends and is distrusted by everyone around him or her, and how this probability would relate to the structure of the network.
	
	\begin{figure}[t]
		\includegraphics[width=\columnwidth]{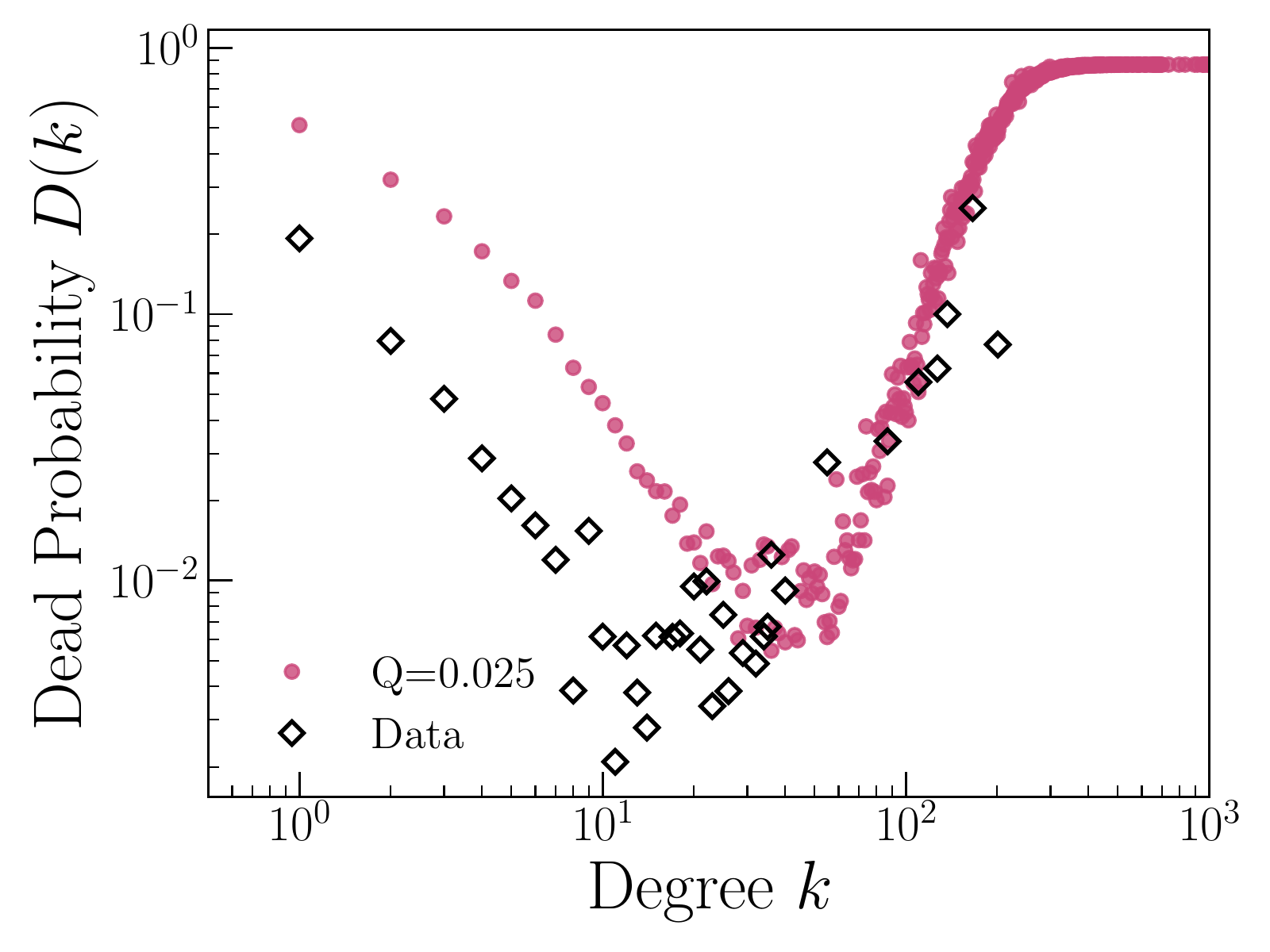}
		\centering	
		\caption{\label{fig:slashdot_sim} Dead probability $D(k)$ obtained from simulation of our link cascade model on the Slashdot network with $Q=0.025$ as compared to the real dataset. For this chosen $Q$ value, our model produces a $D(k)$ curve that is close to the real dataset among the test cases $Q \in \{0.01, 0.02, 0.025, 0.05, 0.075, 0.1 \}$. In particular, the simulated minimum point $k^*_{sim}$ is close to the true $k^*$ with an approximately equal minimal dead probability $D(k^*_{sim}) \approx D(k^*)$.}
	\end{figure}	
	
	We may understand the distribution of these hostile links by modeling their formation and evolution. Since friendship or hostility are inherently link properties, their evolution is naturally described within our framework of link cascade models. Using this language, the probability in question then becomes the dead probability $D(k)$ that a node loses all its friendly links. If one assumes that people do not a priori antagonize each other, all links in the initial network will have sign $+1$. Conflicts may occasionally arise between individuals and turn their linkage from a friendly $+1$ to a hostile $-1$. If this is the full dynamics in the network, then as conflicts occur independently, we should expect a dead probability of the form $D(k) \sim \rho^k$, where $\rho$ is the probability of triggering a conflict. Assuming that $\rho$ is approximately identical across the network, $D(k)$ will monotonically decrease with $k$. However, the effect of conflicts can be more complicated than simply breaking a single friendly link. For instance, if we consider the response of other nearby nodes to a conflict, there could be a possibility of link cascade: After witnessing a conflict outbreak, the nearest neighbors of the original dispute pair may take side on the dispute, and they may choose to agree (keep the $+1$ link) or disagree (change to a $-1$ link) with either side of the pair. As these nearest neighbors reveal their stands, their own neighbors may in turn choose to agree or disagree with them. This process could continue to spread to other neighborhoods and bring more nodes into disagreement. A single conflict in the original dispute can therefore trigger a cascade of conflicts over many links in the network. Such a process is exactly what our link cascade model describes. In this case, we would no longer expect a monotonically decreasing dead probability $D(k)$, but instead a $D(k)$ having a similar form as seen in previous sections.
	
	 Such signed networks have been constructed and studied in Ref.~\onlinecite{leskovec_slashdot}. The network from Ref.~\onlinecite{leskovec_slashdot} is constructed using data collected from the website Slashdot, where users can tag each other as either friends or enemies. The network derived in this study originally takes form of a directed network, but to allow comparison with our proposed link cascade model, we ignore the directionality of the links in the following analysis and treat a link to be negative if it is negative in either direction. We have checked that the result is not affected much by taking into account the link directions (e.g., by considering the in-degree instead of the undirected degree). Table. \ref{table:topology} presents a summary of the network topology, including the number of nodes $N$, number of links $E$, network density $2E / (N(N-1))$, average degree $\langle k \rangle$, degree standard deviation $\sqrt{\langle (k-\langle k\rangle)^2 \rangle}$, and average clustering coefficient $\langle c \rangle$. Fig.~\ref{fig:data} (a) shows the scale-free behavior of the Slashdot network.

	Fig.~\ref{fig:data} (b) shows the dead probability $D(k)$ in the Slashdot network, measured by directly counting nodes that have all negative links for each degree $k$. Strictly speaking, this measurement is only a lower bound of the underlying true dead probability, and the network may not yet reach an equilibrium state. For many $k > 40$ nodes and in fact all nodes with $k > 200$, we observe $D(k) = 0$ from the data (These zeros are not shown in Fig.~\ref{fig:data} because of the logarithmic scale we use). It is possible that there are nodes that are dead in the eventual equilibrium state but have not yet died at the time of observation, and as a result the present measurement based on a single snapshot of the network states underestimates the dead probability. To ensure that it reaches an equilibrium state, detailed time series for the network states would be needed, which however is difficult to obtain for the present study. The Slashdot network from the dataset exhibits a minimum point for $D(k)$, similar to what is observed in our link cascade model. Around the minimum point $k^*$, the data show approximate power law scaling behavior, with $\propto k^{-1.43}$ on the left side and $\propto k^{1.77}$ on the right side. The appearance of a minimum point suggests that the Slashdot friend-enemy network dynamics can be described by the cascading conflict picture we hypothesize rather than being triggered by independent random conflicts.
	
	\begin{figure*}[t]
		\centering
		\includegraphics[width=0.95\textwidth]{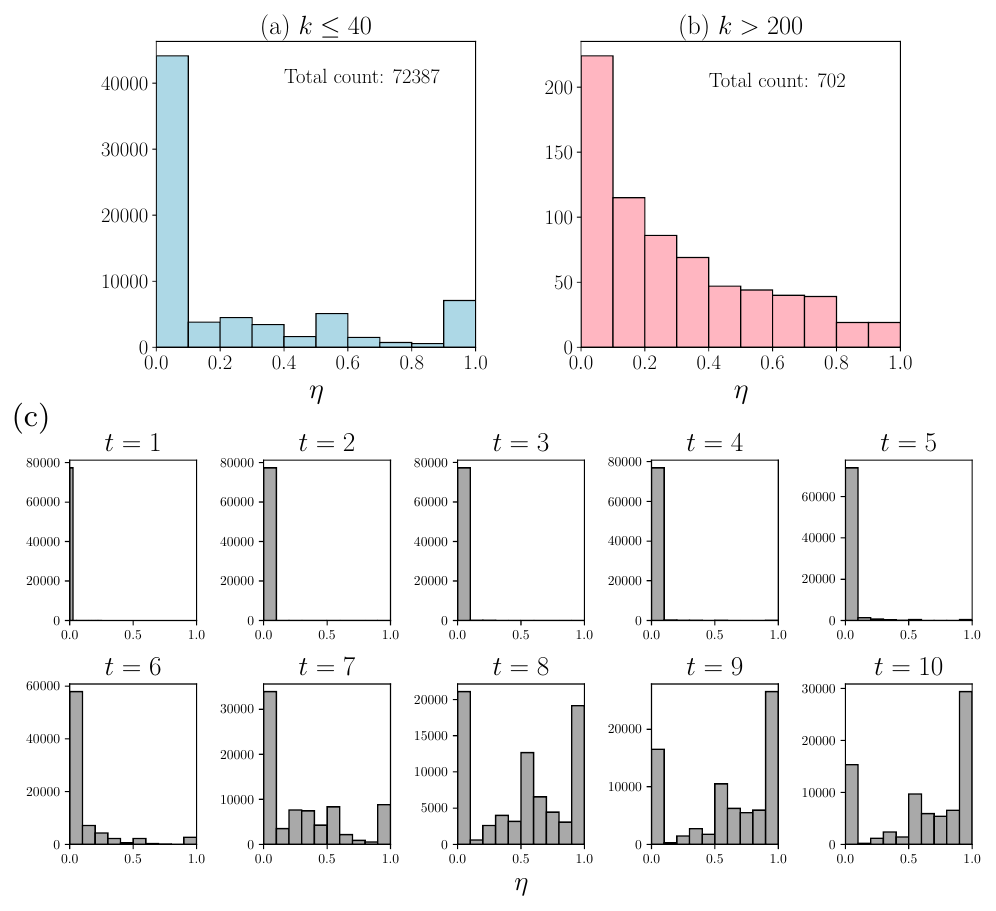}
		\caption{\label{fig:slashdot_neq} Distribution of infected link fraction $\eta$ for (a) $k \leq 40$ nodes and (b) $k > 200$ nodes in the Slashdot network. The simulated time-dependent distribution is plotted in (c). Comparison of (a) and (b) with (c) suggests that both node populations in Slashdot have not yet reached equilibrium. While the shape of (a) is closest to the simulated distribution at $t=7$, the shape of (b) is closet to $t=6$. This suggests that high degree nodes are farther away from equilibrium.}
	\end{figure*}

	To further validate this observation, we perform single-seed simulations of our link cascade model on the Slashdot network topology and compare the final state of the simulations with the measured $D(k)$ from the data. Fig.~\ref{fig:slashdot_sim} shows the simulation results on the Slashdot network with $Q=0.025$. This $Q$ value is found to provide the best-fit curve from the set of test cases $\{0.01, 0.02, 0.025, 0.05, 0.075, 0.1 \}$ that we have performed. Our model is able to reproduce a minimum point $k^*_{sim}$ that is close to the observed $k^*$ and with an approximately equal minimal dead probability $D(k^*_{sim}) \approx D(k^*)$. Given the assumptions of our model, deviation from the data is expected, since: (i) while our model assumed independent cascades, the actual cascade may be correlated. (ii) The diffusion probability $Q$ may vary from link to link and may be time-dependent, which is not considered in our model. In the context of cascading conflict, one expects $Q$ to decay with time as individuals lose interest in the conflict. Such a decaying $Q$ would modify the behavior of $D(k)$ for high $k$, as the cascade may not have enough time to damage all the links in a high $k$ node before $Q$ drops to zero. (iii) While the current study focuses on the single-seed scenario, multiple seeds of cascading damage may be present in reality. Further simulations however show that changing the density of seed link $\rho$ alone does not improve the agreement much between the model and the data. (iv) As mentioned already, the Slashdot network may not yet reach its true equilibrium state. 

	Hints of a non-equilibrium state may be observed by checking the details of negative link distribution against the expected distribution from model simulation. We define the quantity $\eta = n/k$, where $n$ is the number of negative links owned by a node, so that $\eta$ indicates how close a node is to the dead state. Since $\eta$ reflects the progress of infection, the characteristic shape of the distribution $P(\eta)$ across the network will evolve as the link cascade proceeds. We may therefore infer from the measured $P(\eta)$ how closely the Slashdot network is from its equilibrium state. Fig.~\ref{fig:slashdot_neq} presents $P(\eta)$ measured from the Slashdot network from the population of nodes with (a) $k < 40$ and the population with (b) $k > 200$. We separately examine these two node populations since, as we noted above, almost all nodes with $k > 200$ have zero dead probability in the dataset, suggesting that they are far from equilibrium. The distinct shapes of $P(\eta)$ in Fig.~\ref{fig:slashdot_neq} (a) and (b) indicate that the high degree population indeed behaves differently from the low degree population. Fig.~\ref{fig:slashdot_neq} (c) plots the time-dependent $P(\eta)$ distribution for a single realization of our link cascade model on the Slashdot network using the best-fit value $Q=0.025$. Upon comparing Fig.~\ref{fig:slashdot_neq} (a) and (b) with Fig.~\ref{fig:slashdot_neq} (c), we find that $P(\eta)$ measured for both high and low degree nodes resemble the non-equilibrium $P(\eta)$ distribution predicted by our model. While Fig.~\ref{fig:slashdot_neq} (a) closely resembles $t=7$ of Fig.~\ref{fig:slashdot_neq} (c), Fig.~\ref{fig:slashdot_neq} (b) appears closer to $t=6$, This time difference suggests that the high degree nodes are farther away from their true equilibrium states than the low degree nodes, in accord with the observation of zero dead probability of the high degree nodes. From our analysis, one can see that our model could reproduce a family of curves that are close to the real dataset. This suggests that our model has captured a qualitatively correct physical picture of the underlying link cascade process.
	
	We conclude this section by noting that the Slashdot dataset suggests that there is a socially optimal number $k^*$ of connections in order to avoid the node to become a persona non grata. We may interpret this with our earlier combinatorial argument -- having very few friends make one easily isolated but having too many gives too much room for dispute with others. In the Slashdot network, this optimal number $k^*$ is between $15$ and $30$ from our study.

\section{\label{sec:con}Conclusion}

	In this paper, we propose a link independent cascade model as an extension of the original independent cascade model \cite{kempe_IC} in order to explore the consequences of link cascade damage in complex networks. Under link cascade damage, node failures may occur as a result of a node losing all its link connections to the rest of the network. Our model shows that the probability of node failure due to link damage (i.e., the dead probability of a node) $D(k)$ exhibits a nontrivial minimum as a function of node degree $k$. The origin of this minimum lies essentially in the competition between the exponential suppression of probability to infect more links versus the combinatorial growth of infection patterns that infect all the links as $k$ increases. The existence of this minimum point $k^*$ is verified by using numerical simulations for different network topologies and argued via mean-field theory that such a generic feature should present in most networks. As an application of the link cascade model, we apply it to explain the negative link distribution observed in a real-world signed social network by considering the formation of signed social network as a link cascade process. Our model is able to reproduce $D(k)$ curve that are close to real dataset. In particular, the dataset exhibits a minimum as predicted by our model, indicating that our model can adequately capture the dynamics of real-world link cascade processes.

	Although we only consider social applications here, the notion of link damage is general and is relevant as long as the network links are functioning units that are subject to potential failure. As demonstrated in the signed network application, our proposed model serves as a simple working model that can explicate the nature and consequences of link cascade damage. We anticipate our model to have applications in areas such as transportation, biology, and medicine, where link damage is relevant or even prevalent. In this general context, our results regarding dead probability suggest that when link damage is of concern to network design, node strengthening strategies through link addition (e.g., Ref.~\onlinecite{huang_robustness, zeng_robustness, wang_robustness, ji_robustness}) should be performed in such a way that the degree of a node does not deviate too much from $k^*$.

	In the present work, we focus on the equilibrium properties of link cascade. It will be fruitful to extend our current analysis to study degree dependence of link cascade in the non-equilibrium regime. To that end, the mean-field approximation developed in the present work would need to be modified to give a more accurate time-dependent description. Such an extension will shed more light on the nature of link cascade processes and the distinction with its node-based counterpart.

\begin{acknowledgments}
This paper is dedicated to the late Professor K.Y. Szeto, who initiated this research project but unfortunately passed away at the early stage of the present work. He will always be remembered as a mentor and a friend. The authors would also like to thank Chun Yin Yip, Michael K.Y. Wong, and Bradley Foreman for many helpful discussions. K.C. Wong acknowledges support of the Hong Kong PhD Fellowship Scheme (reference number PF19-33123). This research was partially supported by the Research Grants Council of Hong Kong (grant number 16302619).
\end{acknowledgments}

\appendix*
\section{\label{app:min}Minimization of Dead Probability in Star Graph}
	In this appendix, we compute the minimum point of the dead probability in star graph (Eq.~(\ref{eqn:k_star})). Our task is to minimize the function given by Eq.~(\ref{eqn:star_d_expanded}),
	
	\begin{equation}
		\label{aeqn:dk}
		D(k) = Q^{k-1} \frac{(k-1)!}{2 (\ln{2})^{k-1}},
	\end{equation}
	
	\noindent where the domain of $D$ is the set of positive integers. By definition, if $k^*$ is a local minimum of the discrete function $D(k)$, $k^*$ must satisfy the two conditions,

	\begin{subequations}
		\label{aeqn:condition}
	\begin{align}
		\label{aeqn:condition_right}
		\frac{D(k^*)}{D(k^* + 1)} \leq 1, \\
		\label{aeqn:condition_left}
		\frac{D(k^*)}{D(k^* - 1)} \leq 1,
	\end{align}
	\end{subequations}
	
\noindent so that $D(k^*)$ is less than its adjacent values in both directions. The first condition asks for the set of $k$ where $D(k)$ is non-decreasing, and we have

	\begin{equation}
		\frac{D(k)}{D(k+1)} = \frac{\ln{2}}{Q} \left( \frac{1}{k} \right) \equiv \frac{K}{k} \leq 1 \implies k \geq K,
	\end{equation}

\noindent where we have defined the quantity $K = \frac{\ln{2}}{Q}$. The second condition asks for the region where $D(k)$ is non-increasing; solving for $k$ gives

	\begin{equation}
		\frac{D(k)}{D(k-1)} = \frac{k-1}{K} \leq 1 \implies k \leq K + 1.
	\end{equation}
	
\noindent Intersecting the two regions, we found that any integer $k$ within the interval

	\begin{equation}
		K \leq k \leq K+1
	\end{equation}

\noindent can be a solution. We now introduce the ceiling function $f_+$ and the floor function $f_-$, which are defined as

	\begin{align}
		f_+(x) &= \text{Smallest integer $\geq x$}, \\
		f_-(x) &= \text{Largest integer $\leq x$}.
	\end{align}

\noindent We then have, by definition,

	\begin{equation}
		K \leq f_+(K) \leq f_-(K+1) \leq K+1,
	\end{equation}

\noindent and hence both $f_+(K)$ and $f_-(K+1)$ satisfy condition Eq.~(\ref{aeqn:condition}), and they exhaust all possibilities. In fact, because in between $f_+(K)$ and $f_-(K+1)$ there can at most be one integer, we must have $f_+(K) = f_-(K+1)$, unless $K$ is already an integer. On the other hand, if $K$ is already an integer, then by conditions Eq.~(\ref{aeqn:condition}), we have $D(K) = D(K+1)$ and we obtain a degenerate minimum point. In both cases, we reach the conclusion

	\begin{equation}
		k_{star}^* = f_+(K) = f_+ \left( \frac{\ln{2}}{Q} \right) \approx \frac{\ln{2}}{Q}.
	\end{equation}

\section*{Data Availability Statement}
The data that support the findings of this study are available from the corresponding author upon reasonable request.

\bibliography{reference}
\end{document}